\documentclass{aastex631}

\renewcommand{\thefootnote}{\alph{footnote}}

\usepackage{mathrsfs,amssymb}
\usepackage{mathtools}
\usepackage{CJK}
\usepackage{amsmath}
\usepackage{graphicx}
\usepackage{natbib}
\usepackage{epsfig}

\newcommand       \K            {\,{\rm K}}

\newcommand       \s            {\,{\rm s}}

\newcommand       \simali       {\sim\,}

\submitjournal{ApJ}

\begin{document}
\begin{CJK*}{UTF8}{gbsn}

\title{Dependence of Multi-band Absolute Magnitudes and Color Indexes of the Tip of Red Giant Branch Stars on Metallicity in the Galactic Globular Clusters}

\correspondingauthor{Shu Wang, Biwei Jiang, Xiaofeng Wang}
\email{shuwang@nao.cas.cn, bjiang@bnu.edu.cn, wang\_xf@mail.tsinghua.edu.cn}
\author[0000-0002-4448-3679]{Zhenzhen Shao (邵珍珍)}
\affiliation{Institute for Frontiers in Astronomy and Astrophysics,
            Beijing Normal University,  Beijing 102206, China}
\affiliation{School of Physics and Astronomy,
               Beijing Normal University,
               Beijing 100875, China}
\affiliation{Department of Science Research, Beijing Planetarium, Beijing, 100044, China}

\author[0000-0003-4489-9794]{ShuWang (王舒)}
\affiliation{Key Laboratory for Optical Astronomy, National Astronomical Observatories, Chinese Academy of Sciences，Beijing 100101, China}

\author[0000-0003-3168-2617]{Biwei Jiang (姜碧沩)}
\affiliation{Institute for Frontiers in Astronomy and Astrophysics,
            Beijing Normal University,  Beijing 102206, China}
\affiliation{School of Physics and Astronomy,
               Beijing Normal University,
               Beijing 100875, China}

\author[0000-0002-7334-2357]{Xiaofeng Wang (王晓锋)}
\affiliation{Physics Department, Tsinghua University, Beijing 100084, People’s Republic of China}

\author[0000-0002-2614-5959]{Zhishuai Ge (葛志帅)}
\affiliation{Department of Science Research, Beijing Planetarium, Beijing, 100044, China}

\author[0000-0002-7334-2357]{Haichang Zhu (祝海畅)}
\affiliation{Physics Department, Tsinghua University, Beijing 100084, People’s Republic of China}


\begin{abstract}

The tip of red giant branch (TRGB) stars have attracted intensive attention in recent years because their $I$-band absolute magnitudes, $M_{\rm I}$, are often used for distance calibration in the Hubble constant measurements because of its almost independence on metallicity ([Fe/H]). However, a discrepancy exists between various studies and the theoretical stellar model predicts dependence of their luminosity on [Fe/H]. Here we present a careful study of the dependence of absolute magnitudes and color indexes on metallicity in optical and near-infrared bands.
With the TRGB stars identified in 33 Galactic globular clusters by the reddest color in the $G_{\rm BP}-G_{\rm RP}$ vs. $G_{\rm RP}$ diagram, it is confirmed that $M_{\rm I}$ is almost constant of $-4.017 \pm 0.036 \pm 0.027$ mag when $[\rm Fe/H]<-1.2$, which would give $H_0=70.86\pm 1.2\pm0.9$  $\rm kms^{-1} Mp c^{-1}$ with this updated luminosity calibration for type Ia supernovae. However, for $[\rm Fe/H]>-1.2$, $M_{\rm I}$ is found to become fainter with lower metallicity, which would lead to a larger Hubble constant. In the optical $G_{\rm BP}, G_{\rm RP}$ and $V$ bands, the absolute magnitude of TRGB stars tends to increase with metallicity, while in the infrared $J, H$, and $K_{\rm S}$ bands, the variation with metallicity shows an inverse tendency. In addition, the analytical relations of the color indexes with metallicity are presented, which have smaller dispersion than those derived for the corresponding absolute magnitudes.

%

\end{abstract}



\section{Introduction} \label{sec:intro}

The tip of the red giant branch (hereafter TRGB) stars are located at the top of the red giant branch in the Hertzsprung-Russell diagram (hereafter HRD), representing the stage of stellar evolution to the maximum luminosity of red giant branch. The constant $I$-band magnitude of TRGB is widely used as a standard candle to measure the distance of nearby galaxies \citep[][and references therein]{Freedman2019, Hoyt2021}. Recently, TRGB has also been applied to determine the Hubble constant \citep{Freedman2021, Freedman2019, Yuan2019, Anand2021, Soltis2020}, which is a crucial parameter to measure the scale of the universe. There are discrepancies in the Hubble constant measured by different methods. Most notably, the difference between the Hubble constant measured based on Type Ia supernovae (SNe Ia) calibrated by Cepheids and that based on cosmic microwave radiation exceeds five standard deviations (\citealp{riess2021cosmic}; \citealp{collaboration2020planck}), leading to the so-called ``Hubble constant crisis". This discrepancy may be caused by systematic errors in distance measurements or due to unknown new physics. To address the Hubble constant crisis, it is important to find an alternative and reliable luminosity calibrator for SNe Ia. The TRGB calibration provides an independent method of distance measurement, distinct from the Cepheids and cosmological methods, offering a new approach to accurately determine the Hubble constant.

Theoretical foundation supports the use of TRGB as a standard candle for distance measurement.
At the red giant stage, the temperature of the degenerate helium core rapidly rises up to 100 million Kelvin, triggering intense flash of helium fusion (dubbed "helium flash"), which lasts from a few seconds to a few minutes. This rapid evolution causes a discontinuity in the observational characteristics of the stars at the TRGB. Stars with a mass less than 1.8 solar mass have nearly identical degenerate cores during the helium flash, resulting in a nearly constant luminosity at this phase, especially in the $I$ band \citep{Hoyt2021}.
However, analysis of stellar evolution models indicates that stellar parameters such as effective temperature $T_{\rm eff}$, surface gravity $\log g$, metallicity, and bolometric correction affect the conversion of luminosity to absolute magnitudes in different bands \citep{McQuinn2019, Saltas2022}. For example, according to the theoretical isochrones of PARSEC\citep{parsec_Bressan2012,parsec_Marigo2017}, the TRGB magnitudes with a constant age are affected by metallicity in various bands other than the $I$ band. Generally, in the optical bands (e.g. $B, V$), TRGB becomes fainter with metallicity, while in the infrared bands (e.g. $J, K_{\rm S}$), TRGB becomes brighter with metallicity \citep{Freedman2020}.

The TRGB magnitudes in different bands have been studied on the basis of field stars or globular clusters in various galaxies. \citet{Valenti2004b} presented empirical calibrations of the red giant branch (RGB) and TRGB based on 24 Galactic globular clusters and derived the magnitude and metallicity relationships for the RGB and TRGB in the near-infrared $J, H,$ and $K_{\rm S}$ bands.
Later, \citet{Gorski2018} provided empirical calibrations of the absolute magnitude of TRGB in the optical $I$ and near-infrared $J, H,$ $K_{\rm S}$ bands with the $V-K_{\rm S}$, $V-H$, and $J-K_{\rm S}$ colors based on the brightest RGB stars in 19 fields of the Large and Small Magellanic Clouds.
Additionally, \citet{Cerny2020} calibrated the zero point of TRGB in two optical ($V, I$) and three near-infrared ($J, H, K_{\rm S}$) bands based on 46 Galactic globular clusters with [Fe/H]$<-1.0$.
With the available high-precision photometric and spectroscopic data, we have the opportunity to analyze the relationship between multiband magnitudes and colors of TRGB stars with metallicity more comprehensively.

In this work, we used TRGBs in Galactic globular clusters to investigate the variation of absolute magnitude with metallicity.
Globular clusters are ideal objects for TRGB studies because their HRD usually contains the complete evolutionary trajectory of stars, from the main sequence to the red giant branch and the horizontal branch. Furthermore, stars in a globular cluster have nearly identical age and metallicity, which are the two main factors affecting TRGB luminosity. The TRGB is likely to manifest at the tip of the RGB for globular clusters. We identified TRGBs in the Galactic globular clusters using the Gaia DR3 data, which is the third major data release from the Gaia mission, providing accurate and precise astrometric and photometric information for more than a billion stars in our Galactic galaxy \citep{GaiaDR3}. The extensive and highly precise photometric data from Gaia enables an accurate examination of the variation in absolute magnitudes of TRGBs with metallicity.
After cross-matching with the photometric data of 2MASS (Two Micron All Sky Survey, \citealp{2MASS2006}) and OGLE (Optical Gravitational Lensing Experiment, \citealp{OGLE2015}), we study the variation of TRGB absolute magnitude and color with metallicity in optical ($V, I$, $G_{\rm BP}$, $G_{\rm RP}$) and near-infrared ($J, H, K_{\rm S}$) bands.
The structure of this paper is as follows. Section 2 describes the data we use and the method adopted for selecting TRGBs. The relationship of TRGB absolute magnitudes with metallicity and related discussions are presented in Section 3. We summarize our results in Section 4.

\section{Data and Method} \label{sec:data}
To analyze the variation of TRGB absolute magnitudes and colors with metallicity, we first select large globular clusters with different metallicities. Then, we identify TRGBs based on the color-magnitude diagram (CMD) of these globular clusters. The specific process is as follows.

\begin{figure*}
	\vspace{-1mm}
	\centering
	\includegraphics[width={10cm}]{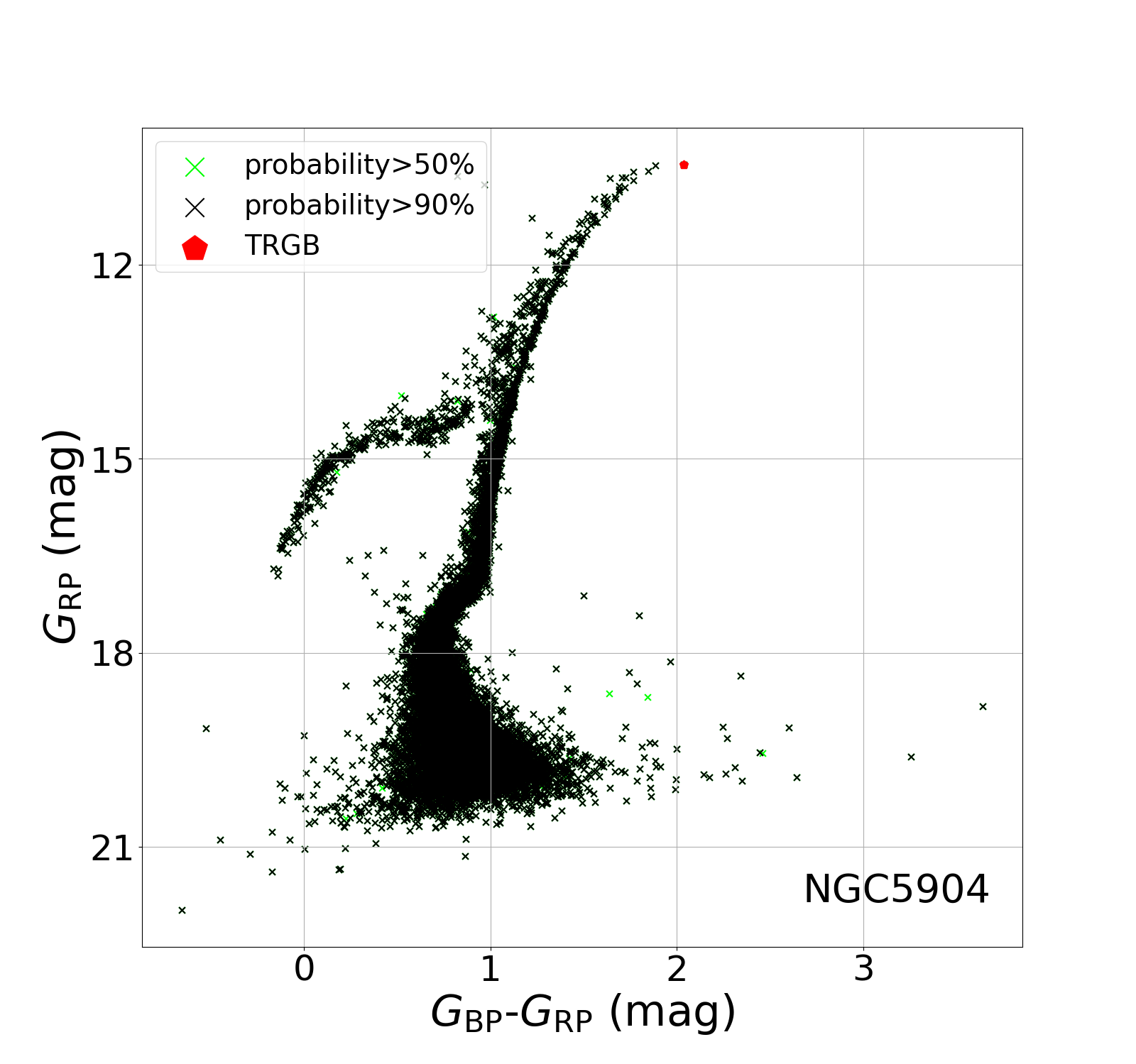}
	
	\caption{
		\label{fig:pmcmd3}
		The color-magnitude diagram of NGC 5904. The green dots represent the stars with over a 50\% probability of belonging to cluster, and the black dots are the selected stars with over a 90\% probability of belonging to the cluster according to \citet{Vasiliev2021Gaia}. The red point is the selected TRGB in this cluster as the reddest star on the red giant branch.
  }
	\vspace{4mm}
\end{figure*}

Considering the scarcity of TRGB stars in globular clusters due to their rapid evolution, we first choose clusters with more than 1000 member stars as our TRGB candidate clusters from \citet{Vasiliev2021Gaia} to ensure the presence of TRGBs. Additionally, clusters that display multiple populations in the $ G_{\rm RP}$ vs. $G_{\rm BP}-G_{\rm RP}$ CMD are removed.
As a result, 43 globular clusters were selected, whose metallicity is uniformly distributed within the range of $-$2.5 to $-$0.5. This reflects that no systematic error is induced in the cluster selection process. 
\citet{Vasiliev2021Gaia} provided detailed GAIA information for 170 Galactic globular clusters, including their positions, proper motions, parallaxes, and number of stars.
They applied a series of stringent quality filters to select stars with reliable astrometric measurements and used these clean subsets to determine the properties of the cluster and the foreground populations, as well as individual membership probabilities for each star, using a mixture modeling procedure detailed in their Section 2 \citep{Vasiliev2021Gaia}. To ensure precision in identifying members of globular clusters, we restrict the probability of stars belonging to the global cluster in \citet{Vasiliev2021Gaia} over 90\% to confirm the association of TRGB with the globular cluster.
This approach allowed us to obtain relatively pure samples of members for these 43 globular clusters. Figure \ref{fig:pmcmd3} shows the distribution of our selected globular cluster members, with NGC 5904 as an example.

Then, we select the TRGB stars from the globular cluster. The widely used method for TRGB detection in the Galactic or extragalaxies is an edge detection algorithm from the discontinuity in the binned luminosity function of bright RGB stars in the CMD (\citealp{Freedman2021, Freedman2019, Anand2021, Hoyt2021}). This method is suitable for galaxies with a large number of stars.  Due to the limited number of stars in globular clusters, the applicability of this method is greatly reduced. \citet{Bellazzini2004} used the edge detection method to study TRGB in globular clusters, but found that only two globular clusters, namely $\omega$ Centauri and 47 Tucanae, are suitable for this method. Therefore, we need to find a new method to detect TRGBs in globular clusters. 

\begin{figure*}
	\vspace{-1mm}
	\centering
	\includegraphics[width={20.6cm}]{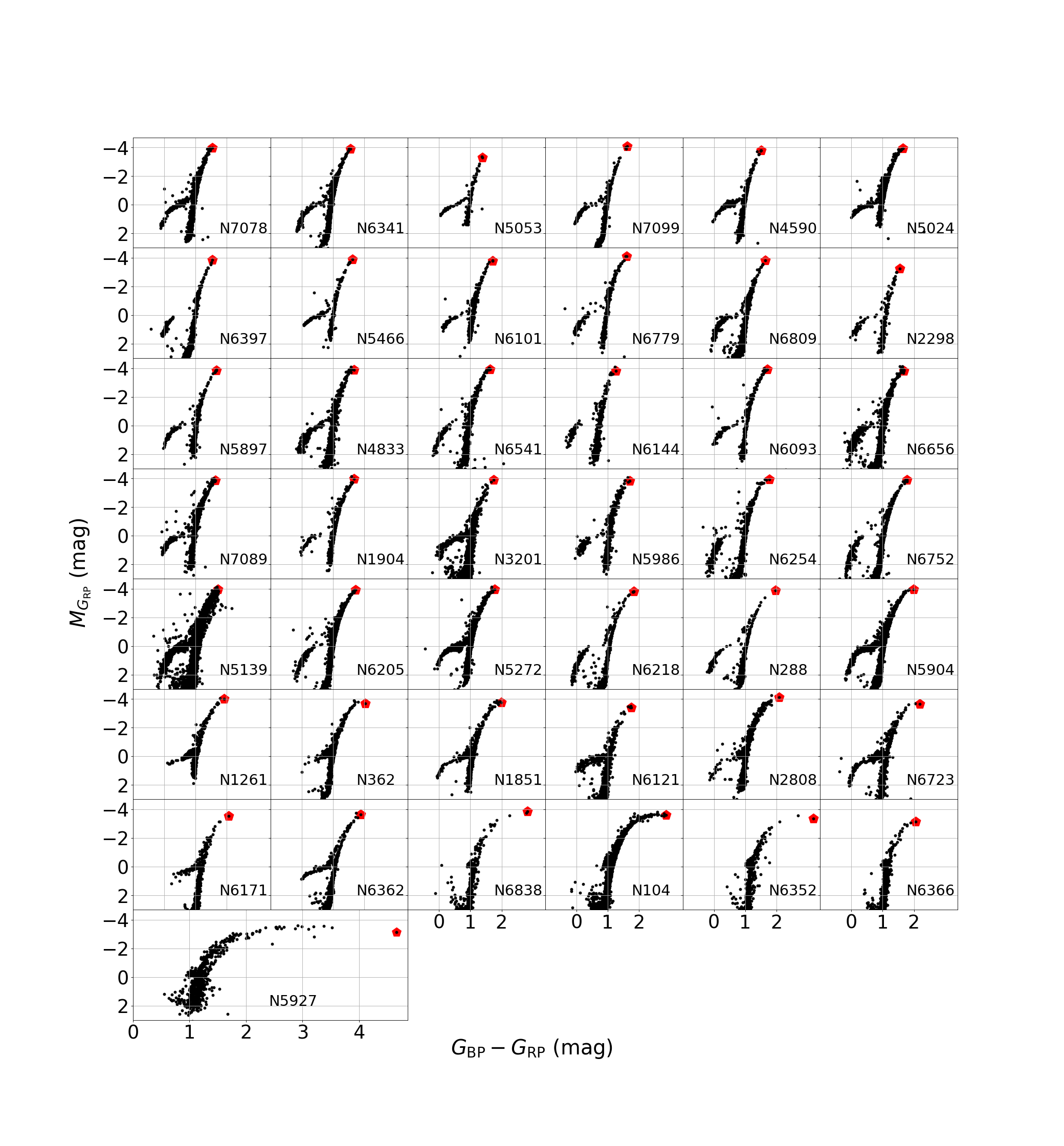}
	
	\caption{
		\label{fig:onepiece}
		The CMD of 43 globular clusters, with the red pentagrams in the maps representing our selected TRGBs.
  }
	\vspace{4mm}
\end{figure*}

In this work, we try to select the reddest stars of the red giant branch from the CMD as TRGB, which is located at the tip of the theoretical evolutionary track of the red-giant branch with the highest luminosity in globular clusters. However, effective temperature, surface gravity, metallicity, and bolometric correction all affect the conversion from luminosity to magnitude \citep{Saltas2022}, so that the highest point of luminosity may not necessarily correspond to the brightest magnitude on some band in the CMD. Particularly in not-so-metal-poor clusters, where opacity is high and only the outer layers of stellar atmospheres are visible, the effective temperature decreases and shifts the blackbody radiation peak towards longer wavelengths. This phenomenon causes a downward bending effect in magnitudes in CMDs, especially at shorter wavelengths \citep{Freedman2020}. Consequently, the TRGB may not be the highest point of the red-giant branch in the CMD but corresponds to the reddest star in globular clusters, assuming constant extinction and distance. Therefore, the TRGB star is identified  as the reddest in the red giant branch of globular clusters, as shown in Figure \ref{fig:pmcmd3} by an asterisk. The CMD of 43 globular clusters and selected TRGBs denoted by a red star are shown in Figure \ref{fig:onepiece}.  It should be mentioned that the reddest TRGB stars may be missed because of the observational limit. The systematic error caused by such an effect will be discussed in Section \ref{sec:3.2uncertainty}.

\begin{figure*}
	\vspace{-1mm}
	\centering
	\includegraphics[width={18cm}]{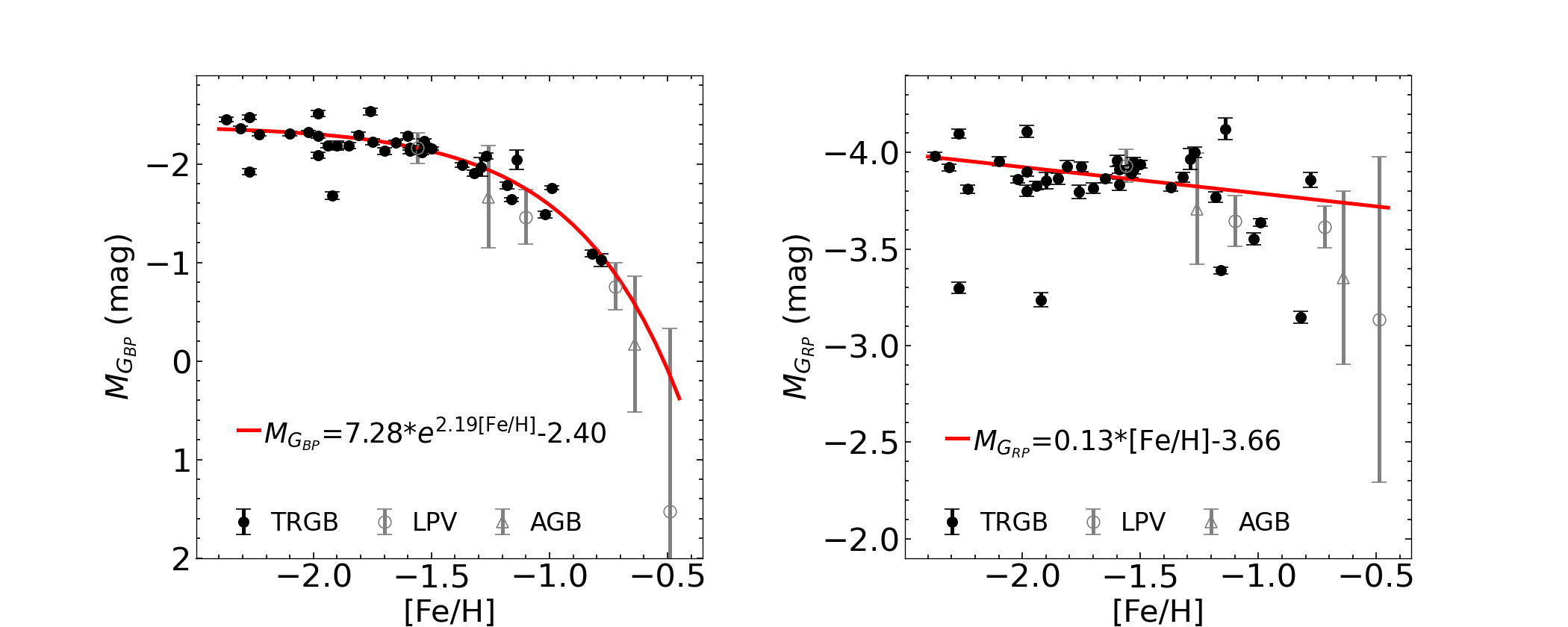}
	
	\caption{
		\label{fig:fitlinesigmalpv2}
		The variation of absolute magnitude in the $G_{\rm BP}$ and $G_{\rm RP}$ band  with metallicity for the 43 TRGB candidates. The red line represents the fitting result, the gray circles and triangle
        indicate LPVs and AGBs, respectively, with error bars representing the photometric dispersion. The black dots represent the remaining TRGBs, and their error bars represent the photometric errors.
	}
	\vspace{4mm}
\end{figure*}

One potential confusion comes from AGB stars that are even redder. To confirm that the stars selected are indeed TRGBs, the 43 TRGB candidate stars are queried in the SIMBAD database, in which 2 are AGB stars and 11 are LPVs \citep{Maiz2023}. As shown in Figure \ref{fig:fitlinesigmalpv2}, green and gray open circles represent LPV and AGB stars, respectively, with error bars representing their photometric dispersions in the Gaia/$G_{\rm BP}$ band. It should be mentioned that the photometric dispersion of Gaia measurements comes from repeated observations of one source and can serve as an indicator of stellar variability. It can be seen that the photometric dispersions of LPVs and AGBs exceed their three-fold photometric errors. According to \citet{Anderson2024}, TRGBs can also be LPVs, which exhibit variations of small amplitude, typically ranging from 0.01 to 0.04 mag. Most of our 11 LPVs fall within this range. Based on the analysis by \citet{Anderson2024} and the measurement errors, a dispersion criterion $<$ 0.05 mag is required as the limit for the amplitude of the TRGB variation, mainly to reduce the uncertainty in photometry.
Consequently, four LPVs (belonging to NGC 104, NGC 5927, NGC 6524, NGC 6723) exceed this limit, so we excluded these sources and two AGB stars, which are marked with an asterisk `*' in Table \ref{table:t1}.

Finally, 37 TRGB stars are selected in the Galactic globular clusters based on the GAIA bands CMD. By cross-matching with the data from 2MASS (\citealp{2MASS2006}) and OGLE (\citealp{OGLE2015}), the TRGB magnitudes are obtained in near-IR $J, H, K_{\rm S}$, and optical $G_{\rm BP}, G_{\rm RP}$, $V, I$ bands. The basic information on these globular clusters are listed in Table \ref{table:t1}.

\section{Results \& Discussions} \label{sec:result}
\subsection{Relationship between Absolute Magnitudes and Metallicity}\label{sec:3.1relation}

With the Gaia parallax and historical distance measurements, \citet{Baumgardt2021} provided distance calibrations for over 160 globular clusters in the Galaxy, including our 37 clusters. With the distance from \citet{Baumgardt2021}, the color excess E(B-V) from \citet{Harris2010}, and the extinction law from \citet{Wang_2019},  the absolute magnitudes of TRGB in the optical and near-infrared bands, i.e. $M_{G_{\rm BP}}$, $M_{G_{\rm RP}}$, $M_J$, $M_{K_{\rm S}}$, $M_V$, $M_I$, are calculated for the 37 globular clusters. Subsequently, the relation between the absolute magnitude of TRGB and the metallicity of the globular clusters of \citet{Harris2010} is determined.

\begin{figure*}
	\vspace{-1mm}
	\centering
	\includegraphics[width={18.2cm}]{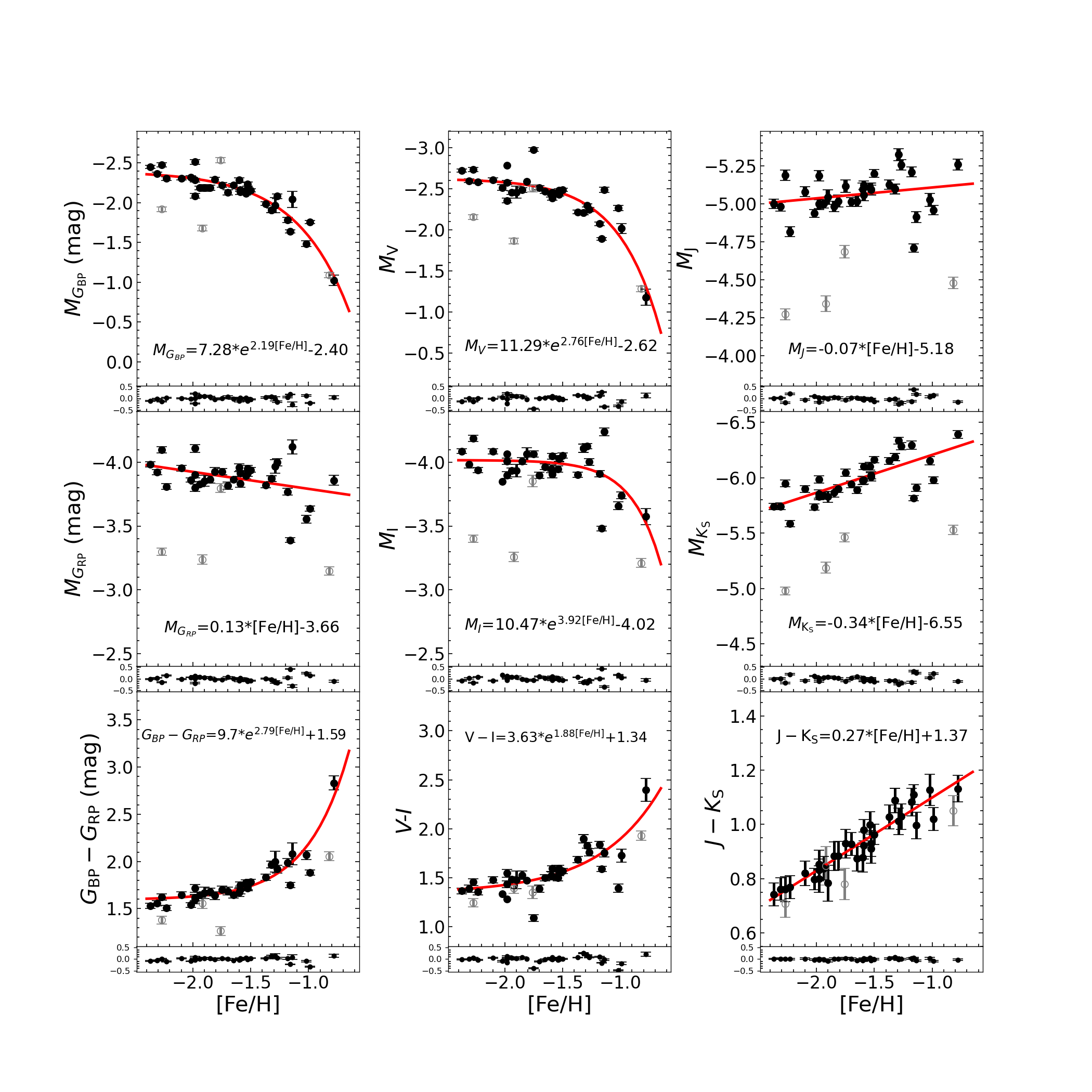}
	
	\caption{
		\label{fig:fehmag}
		The relations of absolute magnitudes or intrinsic color indexes of TRGBs with [Fe/H]. In the $K_{\rm S}$ band, four TRGBs (gray points) deviate from the fitting trend by 3$\sigma$ and are therefore removed in all bands. The red line is the fit of the remaining stars (black points). The distributions of the residuals of the fits are also shown.		
	}
	\vspace{4mm}
\end{figure*}

According to the PARSEC model of stellar evolution (Freedman 2020), the absolute magnitudes of the TRGB vary significantly with metallicity in all bands except for the $I$ band. As shown in Figure \ref{fig:fehmag}, the absolute magnitudes in the $J$ and $K_{\rm S}$ bands, as well as the color index $J-K_{\rm S}$, exhibit a clear linear relationship with metallicity.
In contrast, the absolute magnitudes and color indices in the optical bands, including $G_{\rm BP}, G_{\rm RP}, V$, and $I$, show a dependence on metallicity that more closely resembles an exponential function. To obtain the most suitable fitting form, we applied least-squares fitting to the data for each band using three different functions, i.e. the exponential, the power law, and polynomial. The functional forms and their fitting residuals are summarized in Table \ref{table:t2}.
When the residuals are comparable, the function with fewer parameters is adopted for simplicity. As a result, the absolute magnitudes of the $J$, $K_{\rm S}$, $G_{\rm RP}$ bands, and the color index $J-K_{\rm S}$ are fitted with linear functions, while the absolute magnitudes and color indices of other bands are fitted with exponential functions. It should be noted that these fits are purely mathematical relationships and do not carry any physical significance. 
Figure \ref{fig:fehmag} shows the variation of absolute magnitudes and intrinsic colors of TRGB with metallicity in different bands. The black points represent the selected TRGBs in different clusters, and the red line is the fit of these black points with the fitting formulas also listed in the figure. The distribution of the fitting residuals is also shown. The fitting residual $\sigma$ for the $K_{\rm S}$-band is 0.119. In the $K_{\rm S}$ band panel of Figure \ref{fig:fehmag}, there are four gray points that significantly deviate from the trend line. These outliers fall outside the 3$\sigma$ range of our fitting and are therefore excluded. The fitting in the other bands or intrinsic colors also eliminates these points. The TRGBs of four globular clusters kicked out are marked with a plus sign `+' in Table \ref{table:t1}. Ultimately, 33 TRGB stars were used for the final fitting. The derived relations of multiband absolute magnitudes with metallicity are shown as follows:

\begin{equation}\label{eq:BP}
	M_{G_{\rm BP}}=7.28\times \exp(2.19\times[\rm Fe/H])-2.40 ~~, 
\end{equation}
\begin{equation}
	M_{G_{\rm RP}}=0.13\times [\rm Fe/H]-3.66 ~~,
\end{equation}
\begin{equation}
	M_V=11.29\times \exp(2.76\times[\rm Fe/H])-2.62 ~~,
\end{equation}
\begin{equation}
	M_I=10.47\times \exp(3.92\times[\rm Fe/H])-4.017 ~~,
\end{equation}
\begin{equation}
	M_J=-0.07\times [\rm Fe/H]-5.18 ~~,
\end{equation}
\begin{equation}
	M_{K_{\rm S}}=-0.34\times [\rm Fe/H]-6.55 ~~.
\end{equation}

Obviously, when $[\rm Fe/H]<-1.2$, the absolute magnitude of TRGB in the $I$ band, $M_{\rm I}$, remains nearly constant, hovering around $-4.017 \pm 0.036$ mag, which is consistent with that of $-4.04 \pm 0.015 \pm 0.035$ mag from \citet{Freedman2021}. On the other hand,  $M_{\rm I}$ becomes significantly fainter when $[\rm Fe/H]>-1.2$, with higher metallicity corresponding to a fainter absolute magnitude. A similar trend is shown in the $G_{\rm RP}$ band, which has an effective wavelength close to that of the $I$ band.

In the short-wavelength bands, such as the $V$ and $G_{\rm BP}$ bands, the variation of absolute magnitude ($M_{\rm V}$ and $M_{G_{\rm BP}}$) with metallicity is more significant. $M_{\rm V}$ and $M_{G_{\rm BP}}$ exhibit an almost perfect power-law relationship with metallicity.
In long-wavelength bands, such as the near-infrared band, the absolute magnitude changes with metallicity in the opposite trend, with absolute magnitudes in the $J$ and $K_{\rm S}$ bands ($M_{\rm J}$ and $M_{K_{\rm S}}$) becoming brighter with increasing metallicity. 
As explained theoretically in Section \ref{sec:data}, an increase in the metallicity of TRGB stars leads to higher opacity, which decreases the effective temperature and shifts the peak of the black-body radiation toward longer wavelengths. This shift causes the absolute magnitude of TRGB stars to be brighter in the NIR bands and fainter in the optical bands.
Such variations are consistent with the evolutionary tracks predicted by the stellar evolution theoretical model.

The relation of three color indexes with metallicity is determined independently as following.
\begin{equation}
	{(G_{\rm BP}-G_{\rm RP})}_0=9.7\times \exp(2.79\times[\rm Fe/H])+1.59 ~~,
\end{equation}
\begin{equation}
	{(V-I)}_0=3.63\times \exp(1.88\times[\rm Fe/H])+1.34 ~~,
\end{equation}
\begin{equation}
	{(J-K_{\rm S})}_0=0.27\times[\rm Fe/H]+1.37 ~~.
\end{equation}
It can be seen that the dispersion is smaller for the color index than for the absolute magnitude.

\subsection{Uncertainty Analysis of the Absolute Magnitude of the TRGB} \label{sec:3.2uncertainty}

Distance, color excess, and photometric errors affect the accuracy of the TRGB's absolute magnitude. \citet{Vasiliev2021Gaia} provided the errors in the distances to the Galactic globular clusters. \citet{Legnardi2023} measured the differential reddening of 56 globular clusters, which we approximate as the error in the color excess. These uncertainties were combined with the photometric error to calculate the uncertainty in the absolute magnitude of the TRGB. Focusing on the $I$-band, we derived a mean observational error of 0.027 mag based on the used 33 TRGB stars.

As mentioned in Section \ref{sec:data}, the miss of the reddest TRGB star may introduce a systematic error in the absolute magnitude. To evaluate this effect, we examine the error by taking the second reddest star as TRGB, and find that the absolute magnitude of $I$-band becomes $-3.991$ mag, which is 0.029 mag fainter than that of the reddest TRGB. This value can be regarded as the systematic error caused by source selection. However, as described in Section \ref{sec:3.4}, we have excluded sources that are not considered to be the reddest TRGB candidates. Therefore, finally the selection of TRGB stars should be free of this selection bias.

Combining the absolute magnitude of TRGB in the $I$ band obtained from the fit with its associated uncertainties, we obtain a final value of $M_{\rm I}$ $=-4.017 \pm 0.036 \pm 0.027$ mag. The resultant effect on the Hubble constant will be discussed in Section 3.3.

\subsection{ Comparison with Previous Works}\label{sec:3.3}
So far, many studies focused on the calibration of the absolute magnitude of TRGB in the $I$ band using different sources and methods in the literature, including $M_{\rm I}$ $=-3.97\pm 0.04 \pm 0.1$ mag \citep{Li_2023},  $M_{\rm I}$ $=-3.91\pm 0.05 \pm 0.09$ mag \citep{Li_2022},  $M_{\rm I}$ $=-3.97\pm 0.046$ mag \citep{Yuan2019}, $M_{\rm I}$ $=-4.085 \pm 0.005 \pm 0.1$ mag \citep{Freedman2021}, $M_{\rm I}$ $=-4.042 \pm 0.041 \pm 0.031$ mag \citep{Dixon2023}, with a dispersion of about 0.17 mag. Even for the same object as the LMC, there is a disagreement of 8\% between TRGB calibrations obtained by different methods and sample selections \citep{Hoyt2021}. \citet{Hoyt2021} suggested that this discrepancy could be addressed using a sample selection methodology aimed at filtering out imprecise TRGB signals. In a recent study on the Galactic TRGB using GAIA DR3 data, \citet{Dixon2023} selected Galactic halo stars at high galactic latitudes ($\mid b \mid > 36^{\circ}$) and derived $M_{\rm I}^{\rm TRGB}=-4.042\pm 0.041\pm 0.031$ mag using the Sobel edge detection method. Similarly, \citet{Li_2023} calibrated the TRGB luminosity using GAIA DR3 data and reported $M_{\rm I}^{\rm TRGB}=-3.97\pm 0.04\pm 0.1$ mag using a two-dimensional maximum likelihood algorithm with galactic field stars and Gaia synthetic photometry and parallaxes.

Based on our $M_{\rm I}$-metallicity relation (equation(4)), when $[\rm Fe/H]<-1.2$, $M_{\rm I}$ will be brighter than $-3.92$ mag, which is generally consistent with previous calibrations of TRGB in the Galaxy within the error range.
However, it is worth noting that when $[\rm Fe/H]>-1.2$, $M_{\rm I}$ gradually becomes fainter. For example, for the LMC with higher metallicity ([Fe/H] = -0.9 from \citealt{Liying2024}), the absolute magnitude of the TRGB is measured as $M_{\rm I}$$ = -4.038 \sim -4.047$ mag \citep{Freedman2021,Hoyt2021}, while our prediction is -3.7 mag.
It should be emphasized that the metallicity of our TRGB stars corresponds to that of the LMC galaxy disk. The TRGBs used by \citet{Freedman2021} or \citet{Hoyt2021} were taken from the galactic halo with probably very low metallicity ($< -2$). If assuming metallicity less than -1.6, we predict $M_{\rm I}$ fainter than -4 mag for the TRGB, which is in fact consistent with the results of \citet{Freedman2021} and \citet{Hoyt2021}.

For the LMC, different regions (groups) can exhibit different metallicities and thus different loci of the red giant branch \citep{Choudhury2016,Liying2024}. Previous studies of LMC selected metal-poor regions, which are not suitable for direct comparison with the $M_{\rm I}^{\rm TRGB}$ of average metallicity predicted by our work.
However, the absolute magnitude of TRGB is obviously affected by the metallicity at relatively high metallicity, which needs more attention in the application of TRGB calibration in the future. 
 
\begin{figure*}
	\vspace{-1mm}
	\centering
	\includegraphics[width={8.2cm}]{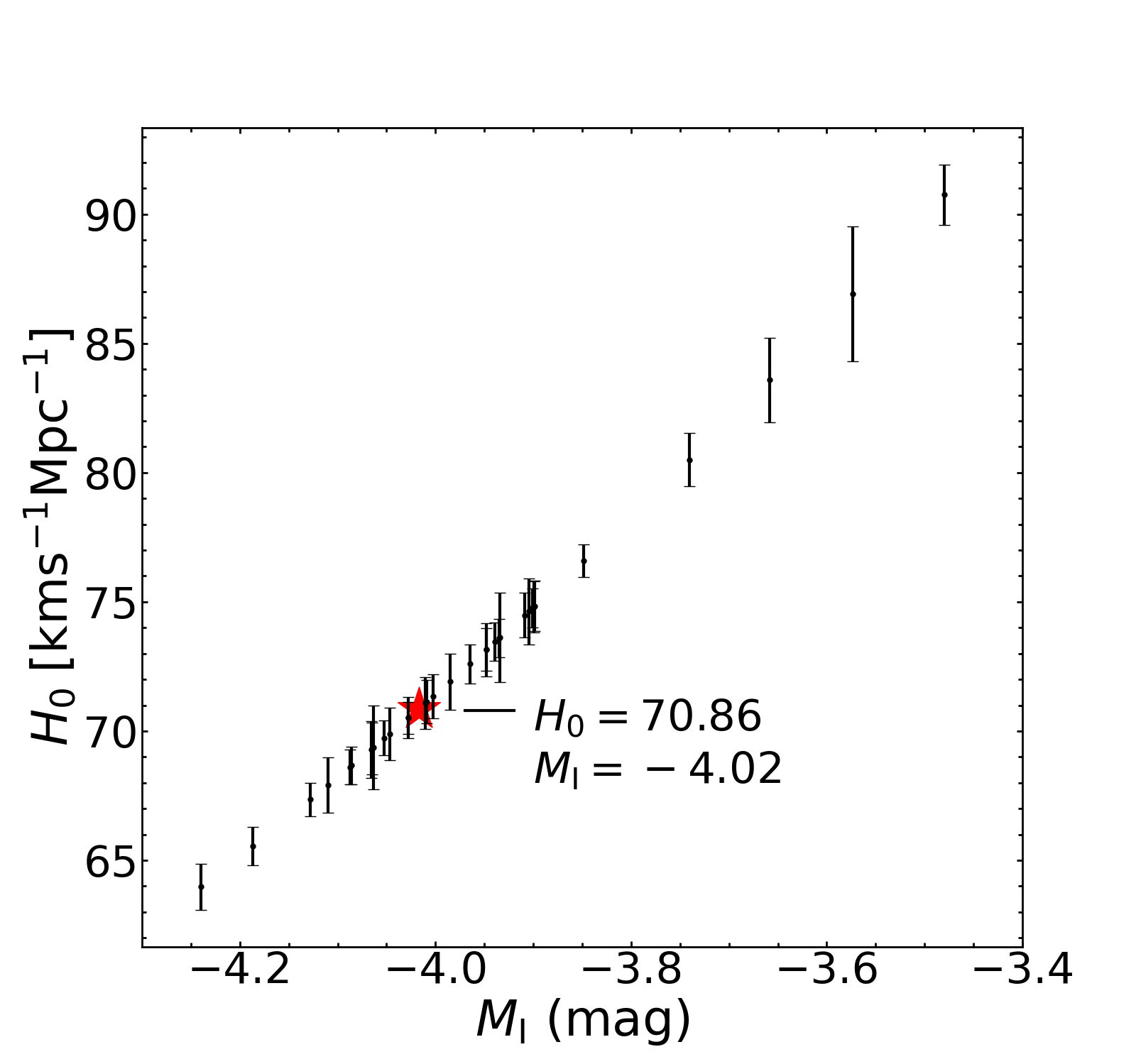}
        \includegraphics[width={8.2cm}]{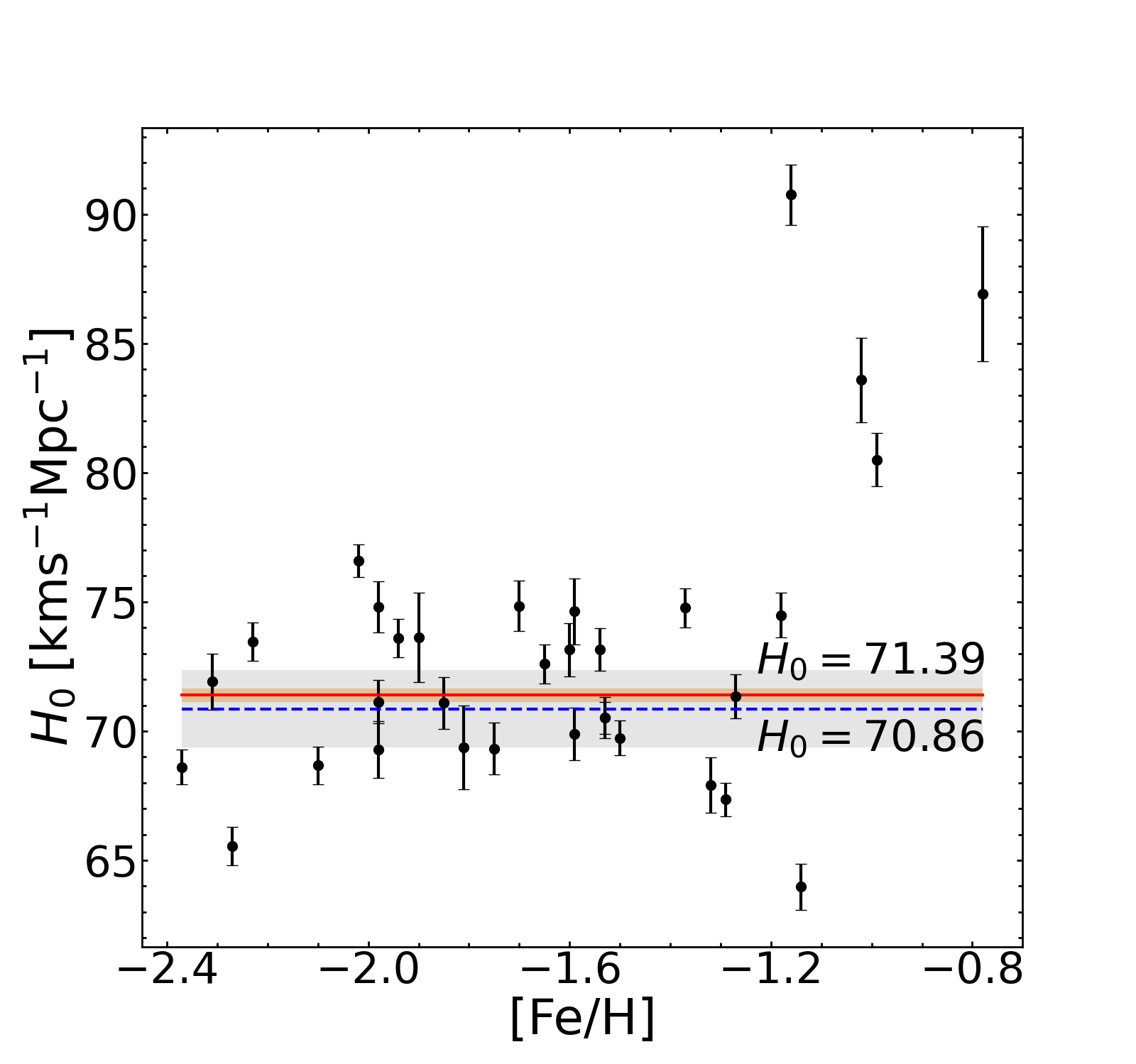}
	
	\caption{
		\label{fig:H0_Imag}
		The relations between Hubble constant ($H_0$) and the $I$-band absolute magnitudes of TRGBs, as well as [Fe/H]. In the left panel, the red pentagram indicates our finalized $H_0$ value and the corresponding [Fe/H]. In the right panel, the red line represents the $H_0$ obtained from fitting sources with metallicity below $-$1.2. The blue dashed line indicates the finalized $H_0$, while the shaded area denotes the associated $1 \sigma$ uncertainty.}	
        
	\vspace{4mm}
\end{figure*}

\citet{Freedman2021} reported the absolute magnitude of TRGB in $I$ band as $M_{\rm I} = -4.049$ mag, which corresponds to Hubble constant $H_0 = 69.8$ $\rm kms^{-1} Mp c^{-1}$. According to Equation 9 ($\log H_0=(M_x^0+5a_x+25)/5$, where x refers to a particular band) in \citet{Riess2016}, changes in $M_{\rm I}$ directly affect $H_0$, and a change of $\sim$ 0.03 mag fainter in $M_{\rm I}$ results in an increase of $H_0$ by 1 $\rm kms^{-1} Mp c^{-1}$. If only $M_{\rm I}$ is varied, the resulting changes in $H_0$ are displayed in the left panel of Figure \ref{fig:H0_Imag}. With our predicted value of $M_{\rm I}^{\rm TRGB}=-4.017 \pm 0.036 \pm 0.027$ mag, the derived Hubble constant $H_0$ is $70.86\pm 1.2\pm0.9$  $\rm kms^{-1} Mp c^{-1}$ \footnote{This error is exclusively propagated from the absolute magnitude uncertainties of the TRGB, without considering the error contribution from supernova.}. If [Fe/H] = -1.5, then $M_{\rm I}$ = -3.99 mag, corresponding to $H_0$ = 71.76 $\rm kms^{-1} Mp c^{-1}$, and if [Fe/H] = -1.2, then $M_{\rm I}$ = -3.972 mag, corresponding to $H_0$ = 72.36 $\rm kms^{-1} Mp c^{-1}$.
The variation of the Hubble constant $H_0$ with metallicity is shown in the right panel of Figure \ref{fig:H0_Imag}. It can be seen that $H_0$ is basically floating around a certain value when [Fe/H] is less than -1.2, and the fit denoted by the red line is about 71.39 $\rm kms^{-1} Mp c^{-1}$. 

From [Fe/H]=-2.0 to [Fe/H]=-1, our derived $M_{\rm I}$ varies by roughly 0.2 mag, while at longer wavelengths like the $K_{\rm S}$ band, it varies by 0.342 mag. Theoretically, \citet{McQuinn2019} used simulated photometry to investigate the dependence of the TRGB luminosity on stellar age and metallicity as a function of wavelength. They found that intrinsic variations in the TRGB magnitude could
increase from a few hundredths magnitude at $0.8<\lambda< 0.9 \mu m$ to approximately 0.6 mag at $\lambda \simali 1.5 \mu m$ as the metallicity increases from -2 to -1. At near-infrared wavelengths redder than in the $I$-band regime, the TRGB becomes brighter as a result of bolometric corrections\footnote{For a thorough explanation of bolometric corrections and their impact on TRGB luminosities, see \citet{Salaris2005}}.

\subsection{Factors Affecting TRGB Luminosity}\label{sec:3.4}
\begin{figure*}
	\vspace{-1mm}
	\centering
	\includegraphics[width={18cm}]{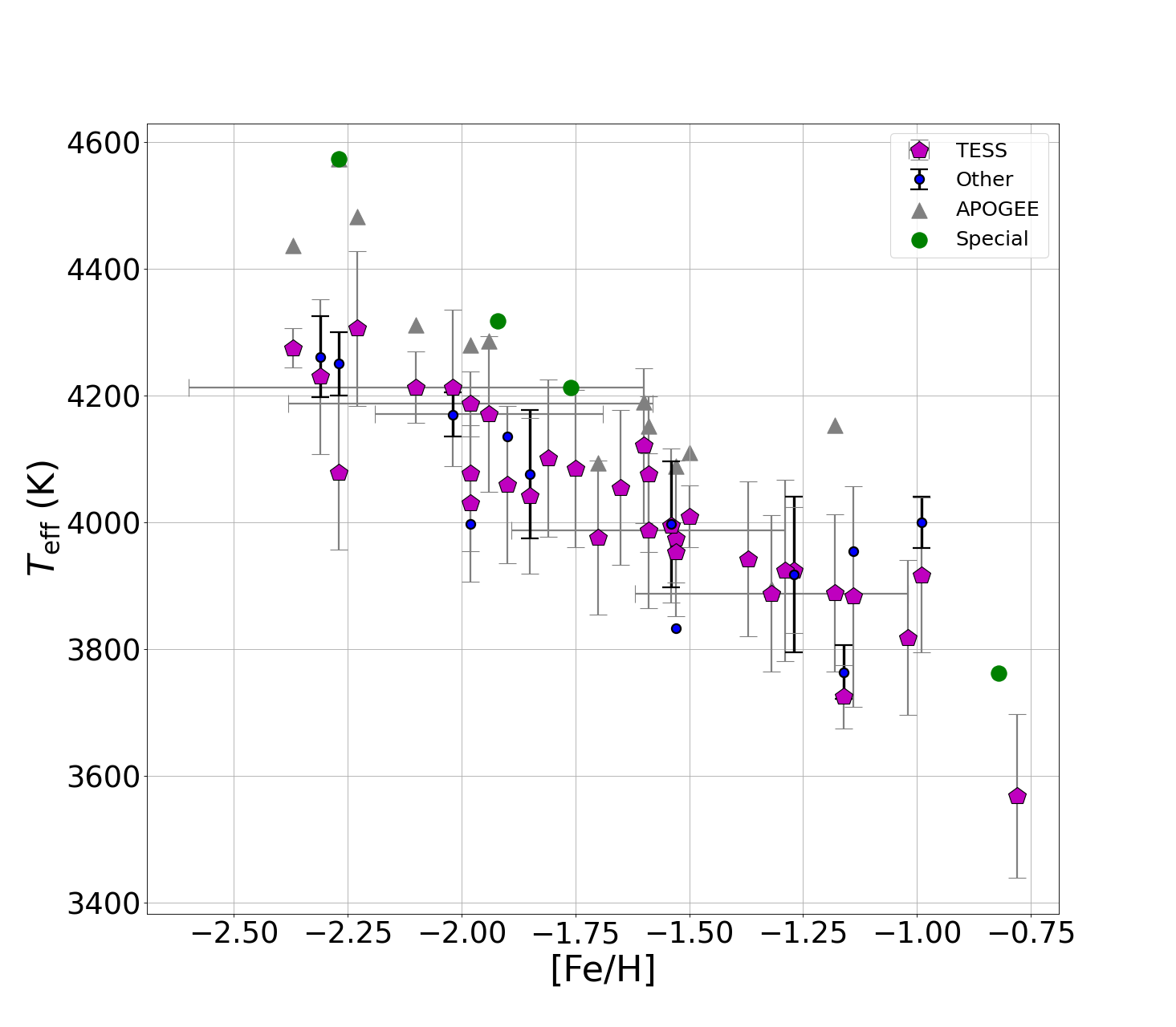}
	
	\caption{
		\label{fig:feh_teff}
		The $T_{\rm eff}$ vs. [Fe/H] diagram of TRGBs. The purple pentagrams, gray triangles, and blue circles are TRGBs with $T_{\rm eff}$ taken from TESS, APOGEE and other observations in the literatures, respectively. The green circles indicate the four excluded sources (see Section \ref{sec:3.1relation} and Figure \ref{fig:fehmag}). The $T_{\rm eff}$ of these four points is generally higher and lie at the top of the trend.
	}
	\vspace{4mm}
\end{figure*}

We have found that the absolute magnitude of the TRGB varies significantly with metallicity.
According to stellar evolution theory, TRGB luminosities are affected by stellar parameters. Metallicity, age, and mass/mass loss are the factors that significantly impact TRGB luminosity \citep{McQuinn2019,Saltas2022}. For the 33 globular clusters, the age is concentrated between 10-13 Gyr \citep{Kruijssen2019}, which has a limited effect on TRGB luminosity. Therefore, metallicity is the primary factor that affects TRGB luminosity.
In addition, extinction and distance can affect the calculation of TRGB absolute magnitude. To determine whether the variation of TRGB magnitude with metallicity is caused by these two factors, we examined the relationship between extinction and distance with metallicity. The extinction $A_\lambda$ was derived from the color excess E(B-V) and the extinction law \citep{Wang_2019}. We found no correlation between either extinction or distance and metallicity. Theoretically, mass loss also has a minor effect on TRGB luminosity. Throughout the evolution from the Main Sequence phase to the TRGB, a star is expected to lose a fraction of its mass, impacting its effective temperature. The effect of mass loss depends on the initial mass and metallicity of the star \citep{Saltas2022}.
To better understand this, the relation between temperatures and metallicity of the TRGB stars is studied in these 37 globular clusters, as shown in Figure \ref{fig:feh_teff}. TRGB temperatures are from the catalog of TESS(Transiting Exoplanet Survey Satellite, \citealp{Tess2019}), APOGEE (Apache Point Observatory Galactic Evolution Experiment, \citealp{Apogeedr17}), and other literatures. Overall, the APOGEE temperatures are higher than those of other observations. The metallicity and temperature of the TRGB show a strong correlation; specifically, higher metallicity corresponds to lower temperature. This indicates that effective temperatures may affect TRGB luminosity, and their effect on luminosity is also reflected in metallicity, since temperature is affected by metallicity. The four points excluded in Figure \ref{fig:fehmag} due to their deviation from the trend line are located at the upper extremity, suggesting that these stars are generally warmer and may not have reached the TRGB stage in their evolution. After this step, the total number of TRGB stars is reduced to 33.

\subsection{Relation between Absolute Magnitudes and Colors}\label{sec:3.5}

\begin{figure*}
	\vspace{-1mm}
	\centering
	\includegraphics[width={18cm}]{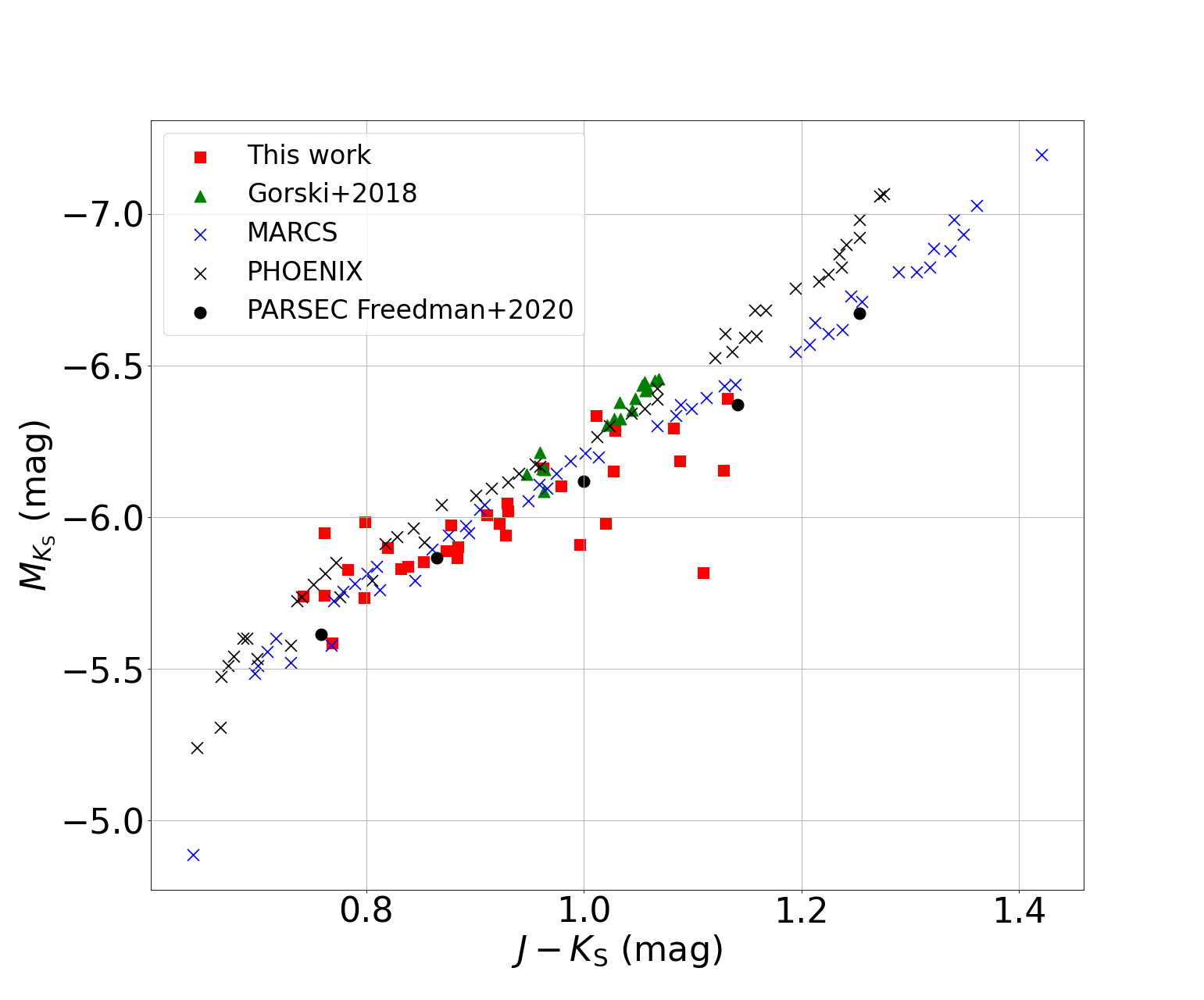}
	
	\caption{
		\label{fig:k_jk}
		The $J-K_S$ vs. $M_{\rm K_S}$ diagram. The TRGB (red square) shows a good linear relation between absolute magnitude and color index, which is in good agreement with the results from various stellar evolution models such as MARCS (blue crosses), PHOENIX (black crosses), PARSEC (black dots), as well as the observations (green triangles) obtained by \citet{Gorski2018} based on the SMC and LMC.
	}
	\vspace{4mm}
\end{figure*}

Previous observations and theoretical studies have found that the TRGB magnitude, especially the $K_{\rm S}$-band magnitude, has a good linear relationship with color. We also examined the variation of TRGB's $K_{\rm S}$-band magnitude with color in 33 globular clusters, as shown in Figure \ref{fig:k_jk}, where the red squares are our results, the blue crosses, black crosses, and black dots represent the theoretical results from stellar evolution models MARCS \citep{MARCS2008Gustafsson}, PHOENIX \citep{Phoenix2008Dotter}, and PARSEC \citep{Parsec2012Bressan, PARSEC2017Marigo} respectively, and the green triangles represent the observations obtained by \citet{Gorski2018} based on the SMC and LMC. Our results agree well with those theoretical models and the observations obtained by \citet{Gorski2018}. There is a TRGB tagged to NGC 6121 that deviates from the trend, and the reason for this deviation is not yet clear. We do not find anything special about this source, and the underlying physical mechanism needs further investigation.

\subsection{Possible Circumstellar Dust in the TRGB}\label{sec:3.6}

\begin{figure*}
	\vspace{-1mm}
	\centering
	\includegraphics[width={18cm}]{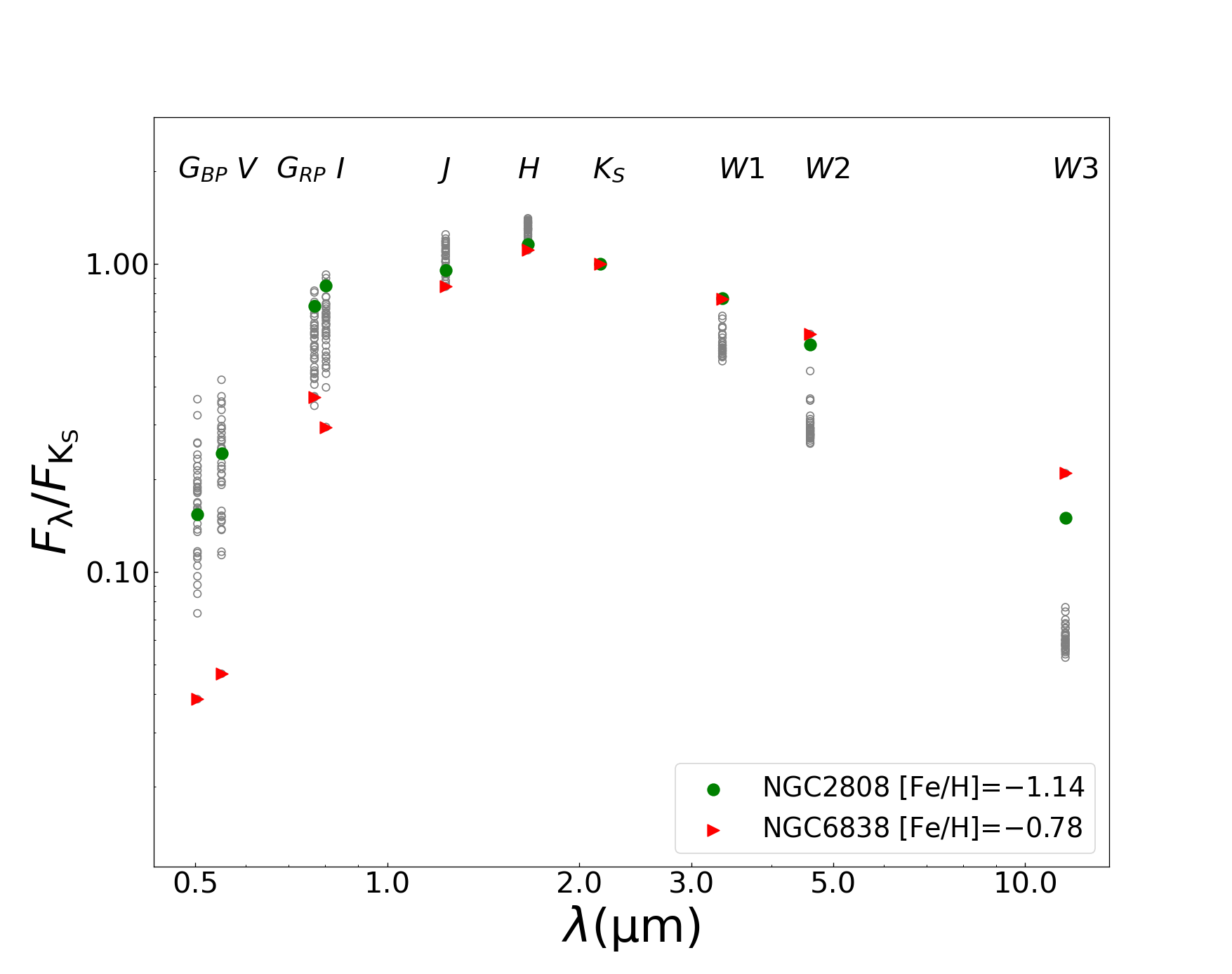}
	
	\caption{
		\label{fig:sed}
  		Spectral energy distributions of the 33 TRGBs. Two TRGBs with infrared excess are represented by different colored symbols, while the remaining stars are indicated by small gray circles.
	}
	\vspace{4mm}
\end{figure*}

Due to mass loss, there may be circumstellar dust around TRGBs. We also checked whether the variation in magnitude of TRGB is due to circumstellar dust.
As mentioned in the previous discussion, $M_{\rm I}$ is significantly fainter when the metallicity is greater than -1.2, which may be caused by the absorption of circumstellar dust. 
To investigate this, we examined the SED of these 33 TRGBs and found that two TRGBs have a significant infrared excess in the mid-infrared bands, such as the $W1$, $W2$, and $W3$ bands in the WISE (Wide-Field Infrared Survey Explorer, \citealp{Wise2012}) survey, as shown in Figure \ref{fig:sed}. This implies that there may exist circumstellar dust around these TRGBs causing extra emission in the infrared bands. The metallicity of these TRGBs is greater than -1.2. We infer that stars with higher metallicity, which are younger and have lower stellar surface temperatures, are more likely to evolve to form circumstellar dust. The fainter TRGB magnitude in the metal-rich stars is most likely caused by circumstellar dust.

Excluding the two TRGBs with infrared excess and refitting the distribution in Figure \ref{fig:fehmag} shows that the trend of TRGB absolute magnitude with metallicity slows down, but this has little effect on the TRGB magnitude in the extremely metal-poor stars. Therefore, in future studies of metal-rich galaxies or clusters, we recommend considering the effect of metallicity on the absolute magnitude of TRGB.

\section{Summary}\label{sec:summary}

Thirty-three globular clusters are selected because of containing over 1000 member stars according to the identification by \citet{Vasiliev2021Gaia} with Gaia astrometric and photometric measurements. Considering different effects of metallicity on various bands, the TRGB star is chosen to be the reddest one of the red-giant branches in the $G_{\rm BP}-G_{\rm RP}$ vs. $G_{\rm RP}$ diagram. After correction for interstellar extinction and distance, the absolute magnitude in the $G_{\rm BP}, V, G_{\rm RP}, I, J, H$, and $K_{\rm S}$ bands is calculated. Their dependence on the metallicity of the globular cluster is then derived. 

The absolute magnitude in the $I$ band, $M_{\rm I}$ is almost constant around $-4.017 \pm 0.036 \pm 0.027$ mag when [Fe/H] is less than -1.2, unaffected by metallicity, which coincides with previous results. This $M_{\rm I}$ corresponds to $H_0=70.86\pm 1.2\pm0.9$  $\rm kms^{-1} Mp c^{-1}$ when it is applied to calibrate the peak luminosity of nearby SNe Ia. However, when [Fe/H] becomes greater than -1.2, $M_{\rm I}$ increases with metallicity. Such a change to fainter magnitude can lead to an increase in the Hubble constant when applied to the cosmological distance calibration. For the optical bands in the $G_{\rm BP}, V, G_{\rm RP}$ band, the absolute magnitude of TRGB becomes fainter with metallicity, while in the near-infrared $J, H, K_{\rm S}$ band, the relation is reversed. This tendency agrees with the stellar theoretical model. In addition, the relation of the color indexes in the optical and infrared bands with metallicity are calculated, which shows apparently smaller dispersion than the absolute magnitude.


\section{Acknowledgment}
\begingroup
\nolinenumbers	

We are grateful to 
X.D. Chen ,Y. Ren, Y. Li and the anonymous referee for their
very helpful comments and suggestions. This work is supported by the National Science Foundation of China (NSFC grants 12288102, 12133002, 12373028, and 12033003), the Beijing Academy of Science and Technology project BGS202205, 24CE-YS-08, National Key R\&D Program of China No. 2019YFA0405500, CMS-CSST-2021-A09, and CMS-CSST-2021-A12. X. W. is also supported by the New Corner Stone Foundation through the Tencent Xplorer Prize. S. W. acknowledge support from the Youth Innovation Promotion Association of the CAS (grant No. 2023065), China Manned Space Program through its Space Application System, National Natural Science Foundation of China under grant No. 12303035, Beijing Natural Science Foundation，No. 1242016. This work has made use of data from GAIA, 2MASS, WISE, OGLE, APOGEE, TESS.

\endgroup


\bibliography{szzref}{}
\bibliographystyle{plain}



\clearpage

%

\clearpage

\begin{footnotesize}
	\begin{longtable}{l|r|r|c|c|c|c|c|c|c|c|c|c}
		\caption{Globular cluster parameters and absolute magnitudes of the cluster's TRGB in different bands.
		}\label{table:t1} \\
		
		\hline
		\hline
		Name & \multicolumn{1}{c|}{RA} & \multicolumn{1}{c|}{DEC} & $\rm Distance^a$ & $E_{(B-V)}^{\rm b}$ &$[\rm Fe/H]^b$&$M_{G_{\rm BP}}$&$M_{\rm V}$&$M_{G_{\rm RP}}$&$M_{\rm I}$&$M_{\rm J}$&$M_{K_{\rm S}}$ &Type\\
		& \multicolumn{1}{c|}{$^{\circ}$} & \multicolumn{1}{c|}{$^{\circ}$}&kpc & mag
		& &mag&mag&mag&mag&mag&mag & \\
		\hline
		\endfirsthead
		\caption{Globular cluster parameters and absolute magnitudes of the cluster's TRGB in different bands.}\\
		\hline
		\hline
	    Name & \multicolumn{1}{c|}{RA} & \multicolumn{1}{c|}{DEC} & $\rm Distance^a$ & $E_{(B-V)}^{\rm b}$ &$[\rm Fe/H]^b$&$M_{G_{\rm BP}}$&$M_{\rm V}$&$M_{G_{\rm RP}}$&$M_{\rm I}$&$M_{\rm J}$&$M_{K_{\rm S}}$ &Type\\
	    & \multicolumn{1}{c|}{$^{\circ}$} & \multicolumn{1}{c|}{$^{\circ}$}&kpc & mag
	    & &mag&mag&mag&mag&mag&mag & \\
		\hline
		\endhead
		\hline
		
		NGC288	&	13.188 	&	-26.583 	&	8.988 	&	0.03 	&	-1.32 	&	-1.905 	&	-2.209 	&	-3.872 	&	-4.110 	&	-5.098 	&	-6.187 & RGB	 \\
		NGC1261	&	48.068 	&	-55.216 	&	16.400 	&	0.01 	&	-1.27 	&	-2.081 	&	-2.242 	&	-4.000 	&	-4.003 	&	-5.258 	&	-6.287 	 & RGB \\
		NGC1851	&	78.528 	&	-40.070 	&	11.951 	&	0.02 	&	-1.18 	&	-1.783 	&	-2.071 	&	-3.769 	&	-3.909 	&	-5.210 	&	-6.293 	 & RGB \\
		NGC1904	&	81.044 	&	-24.524 	&	13.078 	&	0.01 	&	-1.60 	&	-2.285 	&	-2.425 	&	-3.957 	&	-3.948 	&	-5.097 	&	-5.975 	 & RGB \\
		NGC2808	&	138.013 	&	-64.863 	&	10.060 	&	0.22 	&	-1.14 	&	-2.043 	&	-2.485 	&	-4.121 	&	-4.239 	&	-4.914 	&	-5.910  & RGB	 \\
		NGC3201	&	154.403 	&	46.412 	&	4.737 	&	0.24 	&	-1.59 	&	-2.162 	&	-2.454 	&	-3.912 	&	-4.047 	&	-5.123 	&	-6.103  & RGB	 \\
		NGC4590	&	189.867 	&	-26.744 	&	10.404 	&	0.05 	&	-2.23 	&	-2.303 	&	-2.583 	&	-3.810 	&	-3.939 	&	-4.816 	&	-5.585 	 & RGB \\
		NGC4833	&	194.891 	&	-70.877 	&	6.480 	&	0.32 	&	-1.85 	&	-2.187 	&	-2.483 	&	-3.865 	&	-4.010 	&	-4.983 	&	-5.867  & RGB	 \\
		NGC5024	&	198.230 	&	18.168 	&	18.498 	&	0.02 	&	-2.10 	&	-2.307 	&	-2.607 	&	-3.954 	&	-4.086 	&	-5.080 	&	-5.900 	 & RGB \\
		NGC5139	&	201.697 	&	-47.480 	&	5.426 	&	0.12 	&	-1.53 	&	-2.233 	&	-2.481 	&	-3.953 	&	-4.028 	&	-5.097 	&	-6.008 	 & RGB \\
		NGC5272	&	205.548 	&	28.377 	&	10.175 	&	0.01 	&	-1.50 	&	-2.156 	&	-2.484 	&	-3.940 	&	-4.052 	&	-5.200 	&	-6.163 	 & RGB \\
		NGC5466	&	211.364 	&	28.534 	&	16.120 	&	0.00 	&	-1.98 	&	-2.283 	&	-2.574 	&	-3.901 	&	-4.009 	&	-4.999 	&	-5.831 	 & RGB \\
		NGC5904	&	229.638 	&	2.081 	&	7.479 	&	0.03 	&	-1.29 	&	-1.968 	&	-2.297 	&	-3.967 	&	-4.128 	&	-5.324 	&	-6.336  & RGB	 \\
		NGC5986	&	236.512 	&	-37.786 	&	10.540 	&	0.28 	&	-1.59 	&	-2.134 	&	-2.386 	&	-3.834 	&	-3.905 	&	-5.057 	&	-5.979 	 & RGB \\
		NGC6093	&	244.260 	&	-22.976 	&	10.339 	&	0.18 	&	-1.75 	&	-2.224 	&	-2.975 	&	-3.927 	&	-4.065 	&	-5.116 	&	-6.046 	 & RGB \\
		NGC6101	&	246.450 	&	-72.202 	&	14.449 	&	0.05 	&	-1.98 	&	-2.085 	&	-2.354 	&	-3.801 	&	-3.900 	&	-5.000 	&	-5.852 	 & RGB \\
		NGC6121	&	245.897 	&	-26.526 	&	1.851 	&	0.44 	&	-1.16 	&	-1.638 	&	-1.888 	&	-3.389 	&	-3.480 	&	-4.707 	&	-5.817 	 & RGB \\
		NGC6205	&	250.422 	&	36.460 	&	7.419 	&	0.02	&	-1.53 	&	-2.181 	&	-2.431 	&	-3.913 	&	-4.028 	&	-5.091 	&	-6.022 	 & RGB \\
		NGC6341	&	259.281 	&	43.136 	&	8.501 	&	0.02 	&	-2.31 	&	-2.363 	&	-2.595 	&	-3.922 	&	-3.985 	&	-4.982 	&	-5.743 	 & RGB \\
		NGC6397	&	265.175 	&	-53.674 	&	2.482 	&	0.18 	&	-2.02 	&	-2.320 	&	-2.512 	&	-3.860 	&	-3.849 	&	-4.937 	&	-5.735  & RGB	 \\
		NGC6656	&	279.100 	&	-23.905 	&	3.303 	&	0.34 	&	-1.70 	&	-2.130 	&	-2.509 	&	-3.816 	&	-3.899 	&	-5.013 	&	-5.941 	 & RGB \\
		NGC6779	&	289.148 	&	30.183 	&	10.430 	&	0.26 	&	-1.98 	&	-2.512 	&	-2.783 	&	-4.109 	&	-4.066 	&	-5.186 	&	-5.984 	 & RGB \\
		NGC6809	&	294.999 	&	-30.965 	&	5.348 	&	0.08 	&	-1.94 	&	-2.186 	&	-2.452 	&	-3.829 	&	-3.935 	&	-5.000 	&	-5.839 	 & RGB \\
		NGC6838	&	298.444 	&	18.779 	&	4.001 	&	0.25 	&	-0.78 	&	-1.026 	&	-1.178 	&	-3.859 	&	-3.574 	&	-5.259 	&	-6.391 	 & RGB \\
		NGC7078	&	322.493 	&	12.167 	&	10.709 	&	0.10 	&	-2.37 	&	-2.452 	&	-2.721 	&	-3.983 	&	-4.088 	&	-4.998 	&	-5.740   & RGB\\
		NGC7089	&	323.363 	&	-0.823 	&	11.693 	&	0.06 	&	-1.65 	&	-2.220 	&	-2.465 	&	-3.865 	&	-3.964 	&	-5.017 	&	-5.890 	 & RGB \\
		NGC7099	&	325.092 	&	-23.180 	&	8.458 	&	0.03 	&	-2.27 	&	-2.475 	&	-2.731 	&	-4.098 	&	-4.187 	&	-5.188 	&	-5.950 	 & RGB \\
		$\rm NGC2298^+$	&	102.248 	&	-36.005 	&	9.828 	&	0.14 	&	-1.92 	&	-1.680 	&	-1.862 	&	-3.237 	&	-3.256 	&	-4.342 	&	-5.189  & RGB\\
		$\rm NGC5053^+$	&	199.113 	&	17.700 	&	17.537 	&	0.01 	&	-2.27 	&	-1.919 	&	-2.153 	&	-3.299 	&	-3.399 	&	-4.271 	&	-4.979  & RGB\\
		$\rm NGC6144^+$	&	246.808 	&	-26.023 	&	8.151 	&	0.63 	&	-1.76 	&	-2.533 	&	-2.503 	&	-3.797 	&	-3.854 	&	-4.683 	&	-5.463  & RGB\\
		$\rm NGC6366^+$	&	261.934 	&	-5.080 	&	3.444 	&	0.71 	&	-0.82 	&	-1.091 	&	-1.280 	&	-3.147 	&	-3.210 	&	-4.479 	&	-5.529  & RGB\\
		NGC5897	&	229.352 	&	-21.010 	&	12.549 	&	0.09 	&	-1.90 	&	-2.186 	&	-2.455 	&	-3.854 	&	-3.934 	&	-5.045 	&	-5.828 	 & LPV\\
		NGC6101	&	246.450 	&	-72.202 	&	14.449 	&	0.05 	&	-1.98 	&	-2.085 	&	-2.354 	&	-3.801 	&	-3.900 	&	-5.000 	&	-5.852 	& LPV \\
		NGC6171	&	248.133 	&	-13.054 	&	5.631 	&	0.33 	&	-1.02 	&	-1.486 	&	-2.264 	&	-3.553 	&	-3.659 	&	-5.026 	&	-6.154 	& LPV \\
		NGC6218	&	251.809 	&	-1.949 	&	5.109 	&	0.19 	&	-1.37 	&	-1.987 	&	-2.214 	&	-3.821 	&	-3.901 	&	-5.125 	&	-6.152 & LPV	\\
		NGC6362	&	262.979 	&	-67.048 	&	7.649 	&	0.09 	&	-0.99 	&	-1.757 	&	-2.014 	&	-3.637 	&	-3.740 	&	-4.959 	&	-5.979 	& LPV \\
		NGC6541	&	272.010 	&	-43.715 	&	7.609 	&	0.14 	&	-1.81 	&	-2.290 	&	-2.588 	&	-3.928 	&	-4.063 	&	-5.018 	&	-5.902 	& LPV \\
		NGC6752	&	287.717 	&	-59.985 	&	4.125 	&	0.04 	&	-1.54 	&	-2.116 	&	-2.446 	&	-3.892 	&	-3.948 	&	-5.105 	&	-6.104 	& LPV \\
		$\rm NGC104^*$	&	6.024 	&	-72.081 	&	4.521 	&	0.04 	&	-0.72 	&	-0.757 	&	-1.331 	&	-3.615 	&	-3.978 	&	-5.354 	&	-6.547 	& LPV \\
		$\rm NGC5927^*$&	232.003 	&	-50.673 	&	8.270 	&	0.45 	&	-0.49 	&	1.524 	&	1.248 	&	-3.135 	&	-3.399 	&	-6.023 	&	-7.299 	& LPV \\
		$\rm NGC6254^*$	&	254.288 	&	-4.100 	&	5.067 	&	0.28 	&	-1.56 	&	-2.160 	&	-2.566 	&	-3.932 	&	-4.063 	&	-5.152 	&	-6.103 	& LPV \\
		$\rm NGC6723^*$	&	284.888 	&	-36.632 	&	8.267 	&	0.05 	&	-1.10 	&	-1.461 	&	-2.334 	&	-3.645 	&	-3.851 	&	-5.140 	&	-6.209 & LPV	 \\
		$\rm NGC362^*$	&	15.809 	&	-70.849 	&	8.829 	&	0.05 	&	-1.26 	&	-1.666 	&	-2.009 	&	-3.709 	&	-3.705 	&	-5.101 	&	-5.978 	& AGB \\
		$\rm NGC6352^*$	&	261.371 	&	-48.422 	&	5.543 	&	0.22 	&	-0.64 	&	-0.172 	&	-0.797 	&	-3.352 	&	-3.327 	&	-5.381 	&	-6.648 	& AGB \\
		\hline

		\multicolumn{3}{l}{\small $^a$\citet{Vasiliev2021Gaia}} & \multicolumn{2}{l}{\small $^b$\citet{Harris2010} } &
		\multicolumn{4}{l}{\small $^+$ The stars eliminated from fitting }&
		\multicolumn{4}{l}{\small $^*$ The LPV and AGB stars eliminated}
		
		
	\end{longtable}
\end{footnotesize}

\clearpage


\begin{footnotesize}
	\begin{longtable}{l|c|c|c|c|c|c}
		\caption{Fitting residuals of different fitting functions for absolute magnitude variation with metallicity in each band. }\label{table:t2} \\
		
		\hline
		\hline
		Function &$\sigma_{\rm BP}$ & $\sigma_{\rm RP}$ &$\sigma_{\rm V}$& $\sigma_{\rm I}$ & $\sigma_{\rm J}$& $\sigma_{K_{\rm S}}$ \\
		\hline
		\endfirsthead
		\caption{Fitting residuals of different fitting functions for absolute magnitude variation with metallicity in each band.}\\
		\hline
		\hline
		Function &$\sigma_{\rm BP}$ & $\sigma_{\rm RP}$ &$\sigma_{\rm V}$& $\sigma_{\rm I}$ & $\sigma_{\rm J}$& $\sigma_{K_{\rm S}}$ \\
		\hline
		\endhead
		\hline
		$a_0*e^{a_1*{\rm [Fe/H]}}+a_2$	& 0.105 	&	0.129 	&	0.151 	&	0.128 	&	0.126 	&	0.119 \\
		$a_0*a_1^{{\rm [Fe/H]}}+a_2$ &	0.105 	&	0.129 	&	0.151 	&	0.128 	&	0.126 	&	0.119 \\
		$a_0*{\rm [Fe/H]}^3+a_1*{\rm [Fe/H]}^2+a_2*{\rm [Fe/H]}+a_3$	&	0.102 	&	0.128 	&	0.152 	&	0.124 	&	0.125 	&	0.118 \\
		$a_0*{\rm [Fe/H]}^2+a_1*{\rm [Fe/H]}+a_2$	&	0.118 	&	0.128 	&	0.165 	&	0.135 	&	0.125 	&	0.118 \\
		$a_0*{\rm [Fe/H]}+a_1$	& 0.157 	&	0.129 	&	0.199 	&	0.144 	&	0.126 	&	0.119 \\
		
		\hline
		
		
		
	\end{longtable}
\end{footnotesize}

\clearpage

\end{CJK*}	
\end{document}